\documentclass[onecolumn,showpacs]{revtex4}
\usepackage{amssymb}
\usepackage{graphicx}
\usepackage{dcolumn}
\usepackage{bm}

\input epsf

\begin{document}

\title{{\Large Shell Crossing Singularities in Quasi-Spherical Szekeres Models}}

\author{\bf Subenoy Chakraborty$^1$\footnote{subenoyc@yahoo.co.in} and ~Ujjal
Debnath$^2$\footnote{ujjaldebnath@yahoo.com,
ujjal@iucaa.ernet.in}}

\affiliation{$^1$Department of Mathematics, Jadavpur University,
Calcutta-32, India.\\ $^2$Department of Mathematics, Bengal
Engineering and Science University, Shibpur, Howrah-711 103,
India.\\}

\date{\today}

\begin{abstract}
We investigate the occurrence of shell crossing singularities in
quasi-spherical Szekeres dust models with or without a
cosmological constant. We study the conditions for shell crossing
singularity both from physical and geometrical point of view and
they are in agreement.
\end{abstract}

\pacs{0420D, 0420J, 0470B}
\maketitle

\section{{\protect\normalsize \textbf{Introduction }}}

In the study of gravitational collapse, we always encounter with
two types of singularities $-$ shell focusing singularity and
shell crossing singularity. In Tolman-Bondi-Lema\^{\i}tre (TBL)
dust model, these two kinds of singularities will corresponds to
$R=0$ and $R'=0$ respectively. A shell focusing singularity (i.e.,
$R=0$) on a shell of dust occurs when it collapses at or starts
expanding from the centre of matter distribution. The instant at
which a shell at the radial co-ordinate $r$ will reach the centre
of the matter distribution should be a function of $r$. So
different shells of dust arrive at the centre at different times
and there is always a possibility that any two shells of dust
cross each other at a finite radius in course of their collapse.
In this situation the comoving system breaks down, both the
matter density and kretchmann scalar diverge [1,2] and one
encounters the shell crossing singularity. If one treats it as
the boundary surface then the region beyond it is unacceptable
since it has negative density. It is therefore, of interest to
find conditions which
guarantee that no shell crossing will occur.\\

In TBL model, shell crossing singularity has been studied by
several authors [1-9]. Also Goncalves [7] studied the occurrence
of shell crossing in spherical weakly charged dust collapse in
the presence of a non-vanishing cosmological constant. The
positive cosmological constant model conceively prevent the
occurrence of shell crossing thereby allowing at least in
principle for a singularity free `bounce' model. Nolan [8] derive
global weak solutions of Einstein's equations for spherically
symmetric dust-filled space-times which admit shell crossing
singularities. Recently, Hellaby et al [9] investigate the
anisotropic generalization of the wormhole topology in the
Szekeres model. In this work, we have studied the shell crossing
singularity in Szekeres model of the space-time both from
physical and geometrical point of view. In section II, we derive
the basic equations in shell focusing and shell crossing
singularities. We study the physical conditions and geometrical
features of shell crossing singularities in sections III and IV
respectively. Finally the paper ends with a short discussion in
section V.

\section{{\protect\normalsize \textbf{Basic Equations in shell focusing
and shell crossing singularities}}}

Recently, we have presented dust solutions for ($n+2$)-dimensional
Szekeres' space-time model with metric ansatz [10]
\begin{equation}
ds^{2}=dt^{2}-e^{2\alpha }dr^{2}-e^{2\beta }\sum_{i=1}^{n}dx_{i}^{2}
\end{equation}%
where $\alpha $ and $\beta $ are functions of all the ($n+2$)
space-time co-ordinates. If we assume that $\beta ^{\prime
}~(=\frac{\partial \beta }{\partial r})\neq 0$, then the explicit
form of the metric coefficients are

\begin{equation}
e^{\beta}=R(t,r)~e^{\nu(r,x_{1},...,x_{n})}
\end{equation}

and

\begin{equation}
e^{\alpha}=\frac{R^{\prime}+R~\nu^{\prime}}{\sqrt{1+f(r)}}
\end{equation}

where

\begin{equation}
e^{-\nu}=A(r)\sum_{i=1}^{n}x_{i}^{2}+\sum_{i=1}^{n}B_{i}(r)x_{i}+C(r)
\end{equation}

and $R$ satisfied the differential equation

\begin{equation}
\dot{R}^{2}=f(r)+\frac{F(r)}{R^{n-1}}+\frac{2\Lambda}{n(n+1)}~R^{2}~.
\end{equation}

Here $\Lambda $ is the cosmological constant, $f(r)$ and $F(r)$
are arbitrary functions of $r$ alone; and the other arbitrary functions, namely $%
A(r),~B_{i}(r)$'s and $C(r)$ in equation (4) are algebraically
related by the relation

\begin{equation}
\sum_{i=1}^{n}B_{i}^{2}-4AC=-1.
\end{equation}

It is to be noted that the $r$-dependence of these arbitrary
functions $A,~B_{i}$ and $C$ play an important role in
characterizing the geometry of the ($n+1$)-dimensional
space. In fact, the choice $A(r)=C(r)=\frac{1}{2}$ and $B_{i}(r)=0$ (%
$\forall ~i=1,2,...,n$) reduce the space-time metric (1) to the
usual spherically symmetric TBL form

\begin{equation}
ds^{2}=dt^{2}-\frac{R^{\prime }{}^{2}}{1+f(r)}dr^{2}-R^{2}d\Omega
_{n}^{2}.
\end{equation}

by the co-ordinate transformation
\[
\begin{array}{llll}
x_{1}=Sin\theta _{n}Sin\theta _{n-1}...~~...Sin\theta _{2}Cot\frac{1}{2}%
\theta _{1} &  &  &  \\
&  &  &  \\
x_{2}=Cos\theta _{n}Sin\theta _{n-1}...~~...Sin\theta _{2}Cot\frac{1}{2}%
\theta _{1} &  &  &  \\
&  &  &  \\
x_{3}=Cos\theta _{n-1}Sin\theta _{n-2}...~~...Sin\theta _{2}Cot\frac{1}{2}%
\theta _{1} &  &  &  \\
&  &  &  \\
....~~...~~...~~...~~...~~...~~...~~... &  &  &  \\
&  &  &  \\
x_{n-1}=Cos\theta _{3}Sin\theta _{2}Cot\frac{1}{2}\theta _{1} &  &  &  \\
&  &  &  \\
x_{n}=Cos\theta _{2}Cot\frac{1}{2}\theta _{1} &  &  &
\end{array}%
\]%
\newline

Hence in the subsequent discussion we shall restrict ourselves to
the quasi-spherical space-time which is characterized by the $r$
dependence of the function $\nu $ (i.e., $\nu ^{\prime }\neq 0$).
The expression for energy density due to dust matter using the
Einstein equations is

\begin{equation}
\rho(t,r,x_{1},...,x_{n}) =\frac{n}{2}~\frac{F^{\prime }+(n+1)F\nu ^{\prime }%
}{R^{n}(R^{\prime }+R\nu ^{\prime })}.
\end{equation}

The space-time singularity will occur when either (i) $R=0$ i.e.,
$\beta =-\infty $ or (ii) $R'+R\nu'=0$ i.e., $\alpha =-\infty $.
The standard terminology for spherical collapse suggests that the
first case corresponds to shell focusing singularity while in the
second case we have a shell-crossing singularity. In the
following we shall discuss the situations for shell crossing
singularity.
\newline

Suppose the collapse develops at the initial hypersurface
$t=t_{i}$ where we assume $R(t_{i},r)$ to be a monotonically
increasing function of $r$. So, without any loss of generality,
we can label the dust shells by the choice $R(t_{i},r)=r$. Hence
the expression for the initial density distribution is given by

\begin{equation}
\rho_{i}(r,x_{1},...,x_{n})=\rho(t_{i},r,x_{1},...,x_{n})=\frac{n}{2}~\frac{%
F^{\prime}+(n+1)F\nu^{\prime}}{r^{n}(1+r\nu^{\prime})}
\end{equation}

If \ we consider that the collapsing process starts from a regular
initial hypersurface then the function $\rho _{i}$ must be
non-singular (and also positive from physical point of view).
Moreover the local flatness property of the space-time near $r=0$
demands $f(r)\rightarrow 0$ as $r\rightarrow 0$. Then in order to
$\dot{R}^{2}$ to be bounded as $r\rightarrow 0$ we must have
$F(r)\sim O(r^{m})$ where $m\geq n-1$ (see eq. (5)). On the other hand, for small $r$, $%
\rho _{i}(r)\simeq \frac{n}{2}\frac{F^{\prime
}+(n+1)F\nu^{\prime}}{r^{n}}$ and consequently, for regular $\rho
_{i}(r)$ near $r=0$, we must have $F(r)\sim
O(r^{n+1})$ and $\nu'\sim O(\frac{1}{r})$. Hence, starting from a regular initial hypersurface, we can express $F(r)$ and $%
\rho _{i}(r)$ as a power series near $r=0$ as [11]

\begin{equation}
F(r)=\sum_{j=0}^{\infty }F_{j}~r^{n+j+1}
\end{equation}%
and
\begin{equation}
\rho _{i}(r)=\sum_{j=0}^{\infty }\rho _{j}~r^{j}.
\end{equation}

As $\nu ^{\prime }$ appears in the expression for the density as
well as in the metric coefficient, so we can write [11]

\begin{equation}
\nu ^{\prime }(r)=\sum_{j=-1}^{\infty }\nu _{j}~r^{j}
\end{equation}%
where $\nu _{_{-1}}>-1$.\newline

Now, using these series expansions in equation (9) we have the following
relations between the coefficients,

$$
\rho_{0}=\frac{n(n+1)}{2}F_{0},~~\rho_{1}=\frac{n}{2}\left(n+1+\frac{1}{1+
\nu_{_{-1}}}\right)F_{1},
$$
\vspace{-5mm}

$$
\rho_{2}=\frac{n}{2}\left[\left(n+1+\frac{2}{1+ \nu_{_{-1}}}\right)F_{2}-%
\frac{F_{1}\nu_{_{0}}}{(1+\nu_{_{-1}})^{2}}\right],
$$
\vspace{-5mm}

$$
\rho _{3}=\frac{n}{2}\left[ \left( n+1+\frac{3}{1+\nu _{_{-1}}}\right) F_{3}-%
\frac{2F_{2}\nu _{_{0}}}{(1+\nu _{_{-1}})^{2}}-\frac{(1+\nu
_{_{-1}})\nu _{_{1}}-\nu _{_{0}}^{2}}{(1+\nu
_{_{-1}})^{3}}F_{1}\right]
$$%
and so on.\newline

Now in order to form a singularity from the gravitational collapse
of dust, all portions of the dust cloud should collapse i.e., $\dot{%
R}\leq 0$. Let us denote by $t_{sf}(r)$ and $t_{sc}(r)$ as the
time for shell-focusing and shell-crossing singularities
occurring at radial coordinate $r$. Hence we have the relations
\begin{equation}
R(t_{sf},r)=0
\end{equation}%
and

\begin{equation}
R^{\prime }(t_{sc},r)+R(t_{sc},r)\nu ^{\prime }(r,x_{1},x_{2},...,x_{n})=0.
\end{equation}%
Note that `$t_{sc}$' may also depend on $x_{1},x_{2},...,x_{n}$.\newline

\section{{\protect\normalsize \textbf{Physical conditions for shell crossing singularity}}}

We shall now make a comparative study of shell focusing and shell
crossing singularity time and find conditions in favour (or
against) of formation of shell crossing singularity for the
following different choices:\newline

(i)~~ $f(r)=0$, $\Lambda= 0$:\newline

In this case equation (5) can be integrated to give

\begin{equation}
R^{\frac{n+1}{2}}=r^{\frac{n+1}{2}}-\frac{n+1}{2}\sqrt{F(r)}~(t-t_{i})
\end{equation}

So $R(t_{sf},r)=0$ results

\begin{equation}
t_{sf}(r)=t_{i}+\frac{2}{(n+1)\sqrt{F(r)}}~r^{\frac{n+1}{2}}
\end{equation}

Now to avoid the shell crossing singularity either all shells
will collapse at the same time (i.e., $t_{sf}$ is independent of
$r$) or larger shell will collapse at late time (i.e.,
$t_{sf}(r)$ is a monotone increasing function of $r$). These two
conditions can be combined as
$$
t'_{sf}(r)\ge 0
$$

or equivalently from equation (16)

\begin{equation}
\frac{F'(r)}{F(r)}\le \frac{n+1}{r}
\end{equation}

Now combining equations (14) and (15) we have

\begin{equation}
t_{sc}(r)-t_{sf}(r)=\frac{2r^{\frac{n+1}{2}}\left\{\frac{n+1}{r}-\frac{F'(r)}{F(r)}
\right\}} {(n+1)\sqrt{F(r)}\left\{\frac{F'(r)}{F(r)}+(n+1)\nu'
\right\}}
\end{equation}

But if it is so happen that $R'+R\nu'=0$ is a regular extremum for
$\beta$, then we must have finite $\rho$. This implies from
equation (8) that $F'+(n+1)F\nu'=0$. Hence, if there is no shell
crossing singularity corresponding to equation (14) we must have
two
possibilities:\\

\begin{equation}
\text{either~~~~(a)}~~\frac{F'(r)}{F(r)}+(n+1)\nu'=0~~~~\text{and}~~~
\frac{n+1}{r}-\frac{F'(r)}{F(r)}=0
\end{equation}

\begin{equation}
\text{or~~~~~~~~(b)}~~\frac{F'(r)}{F(r)}+(n+1)\nu'=0~~~~\text{and}~~~
\frac{n+1}{r}-\frac{F'(r)}{F(r)}>0
\end{equation}

For the first choice $t_{sf}$ is constant, so all shells collapse
simultaneously while for the second choice $t_{sf}$ is a monotonic
increasing function of $r$ and there is an infinite time
difference between the occurrence of both type of singularities.\\

The value of $R$ at $t=t_{sc}(r)$ is

\begin{equation}
\left\{R(t_{sc},r)\right\}^{\frac{n+1}{2}}=\frac{r^{\frac{n+1}{2}}
\left\{\frac{F'(r)}{F(r)}-\frac{n+1}{r}\right\}}
{\left\{\frac{F'(r)}{F(r)}+(n+1)\nu' \right\}}
\end{equation}

Therefore as an complementary event, the conditions for occurrence
of shell crossing singularity are
$R'+R\nu'=0,~\rho=\infty,~\dot{R}<0,~R>0,~t'_{sf}<0$.\\

As $t'_{sf}<0$ implies

\begin{equation}
\frac{F'(r)}{F(r)}>\frac{n+1}{r}
\end{equation}

so $R>0$ demands

\begin{equation}
\frac{F'(r)}{F(r)}+(n+1)\nu'>0
\end{equation}

Hence we have
\begin{equation}
F(r)\sim r^{l} ~~\text{and}~~ e^{\nu}\sim r^{p}
\end{equation}

for shell crossing singularity with $l>(n+1)$ and $(n+1)p>-l$.\\
\\

(ii)~~ $f(r)=0$, $\Lambda\ne 0$:\newline

This choice will give the solution to equation (5) as [11]

\begin{equation}
t=t_{i}+\sqrt{\frac{2n}{(n+1)\Lambda}}\left[Sinh^{-1}\left(\sqrt{\frac{%
2\Lambda r^{n+1}}{n(n+1)F(r)}}\right)-Sinh^{-1}\left(\sqrt{\frac{2\Lambda
R^{n+1}} {n(n+1)F(r)}}\right)\right]
\end{equation}%
At the shell focusing time $t_{sf}(r)$,~$R=0$, hence we have

\begin{equation}
t_{sf}(r)=t_{i}+\sqrt{\frac{2n}{(n+1)\Lambda}}~Sinh^{-1}\left(\sqrt{\frac{%
2\Lambda r^{n+1}}{n(n+1)F(r)}}\right)
\end{equation}

Since to avoid the shell crossing singularity, $t_{sf}(r)$ should
be an increasing (or a constant) function of time i.e.,
$t'_{sf}(r)\ge 0$, which implies the same condition (17) as in
case (i) and is independent of $\Lambda$ (whether zero or not).
Further, substitution of equation (25) in equation (14) will give
\begin{equation}
t_{sc}(r)-t_{sf}(r)=tanh^{-1}\left[\frac{\sqrt{\frac{2(n+1)\Lambda}{n}}
~t'_{sf}(r)}{\left\{\frac{F'(r)}{F(r)}+(n+1)\nu'\right\}}\right]
\end{equation}

But if there is no shell crossing singularity corresponding to
$R'+R\nu'=0$ (then it will correspond to an extremum of $\beta$)
then $\rho$ must be finite. This will be possible only when
$F'+(n+1)F\nu'=0$. But from eq.(27) it is permissible only when
$t'_{sf}=0$ i.e., $t_{sf}$ is independent of $r$.\\\\

(iii)~~ $f(r)\neq 0, \Lambda=0,~ \dot{R}(t_{i},r)=0$ (\textit{time
symmetry}):\newline

In this case explicit solution is possible only for five
dimension (i.e., for $n=3$) and the result as
\begin{equation}
R^{2}=r^{2}-\frac{F(r)}{r^{2}}(t-t_{i})^{2},
\end{equation}%
But the shell focusing condition $R(t_{sf},r)=0$ gives
$$
t_{sf}=t_{i}+\frac{r^{2}}{\sqrt{F(r)}}.
$$

So $t'_{sf}\ge 0$ will give
\begin{equation}
\frac{F'}{F}\le \frac{4}{r}
\end{equation}

Here the time difference between the two types of singularities is
\begin{equation}
t_{sc}-t_{sf}=\frac{r^{2}}{\sqrt{F}}\left[\sqrt{\frac{\nu'+\frac{1}{r}}
{\frac{F'}{2F}+\nu'-\frac{1}{r}}}~~-1 \right]
\end{equation}

The r.h.s. of equation (30) always positive by the inequality
(29).\\

\section{{\protect\normalsize \textbf{Geometrical features of shell crossing singularity}}}

Now we shall discuss the shell crossing singularity from
geometrical point of view. We note that $R'+R\nu'$ (related to
$e^{\alpha}$) is always $\ge 0$. The equality sign corresponds to
the shell crossing singularity. In fact geometrically, a shell
crossing singularity (if it exists) is the locus of zeros of the
function $R'+R\nu'$ (i.e., $\alpha=-\infty$). Now writing
explicitly the function $R'+R\nu'$ using the solution (4) for
$\nu$ we have

\begin{equation}
R'+R\nu'=e^{\nu}\left[(R'A-RA')\sum_{i=1}^{n}x_{i}^{2}+\sum_{i=1}^{n}(R'B_{i}-RB'_{i})~x_{i}+
(CR'-C'R) \right]
\end{equation}

We see that (a detailed analysis is given in the $\bf Appendix$)
there will be no shell crossing singularity i.e., $R'+R\nu'$ will
be positive definite if

\begin{equation}
\frac{R'^{2}}{R^{2}}>\sum_{i=1}^{n}B'^{2}_{i}-4A'C'=\psi(r)~\text{(say)}
\end{equation}

Note that $R'+R\nu'$ will also be positive for
$\frac{R'^{2}}{R^{2}}=\psi(r)$, provided $x_{i}\ne
\frac{RB'_{i}-R'B_{i}}{2(R'A-RA')}=x_{0i}~,~i=1,2,...,n$. Thus
shell crossing is a single point $(x_{01},~x_{02},...,x_{0n})$ in
the constant ($t, r$)-hypersurface ($n$ dimensional). In other
worlds, it is a curve in the $t$-constant ($n+1$)-D hypersurface
and a 2 surface in ($n+2$)-D space-time.\\

When $\frac{R'^{2}}{R^{2}}<\psi(r)$ then shell crossing
singularity lies on $n$-hypersphere in the $n$-dimensional
$x_{i}$'s plane. This hypersphere has centre
$(x_{01},~x_{02},...,x_{0n})$ and radius

\begin{equation}
r_{c}=\frac{\sqrt{R^{2}\left(\sum_{i=1}^{n}B_{i}'^{2}-4A'C'\right)-R'^{2}}}{2(R'A-RA')}
\end{equation}

In the above we have assumed $a=(R'A-RA')$ to be positive.
However, if $a<0$ and $\frac{R'^{2}}{R^{2}}<\psi(r)$ then also
shell crossing singularity is possible and it lies on an
$n$-hypersphere having same centre $(x_{01},~x_{02},...,x_{0n})$
and radius $r_{c}$. Further, there will be no shell crossing
singularity if the variables $x_{i}$'s lie inside the above
$n$-hypersphere.\\

The above hypersphere is different from the hypersphere with
$\nu'=0$ i.e.,

\begin{equation}
A'(r)\sum_{i=1}^{n}x_{i}^{2}+\sum_{i=1}^{n}B'_{i}(r)x_{i}+C'(r)=0
\end{equation}

So the shell crossing set intersects with the surface of constant
$r$ and $t$ along the line (curve) $\frac{R'}{R}=-\nu'$=constant.\\

Now for positive density we note that $F'+(n+1)F\nu'$ and
$R'+R\nu'$ must have the same sign. We now consider the case
where both are positive (when both are negative, we just reverse
the inequalities). When both are zero then it can happen for a
particular value of $x_{i}$'s ($i=1,2,...,n$) if
$\frac{F'}{(n+1)F}=\frac{R'}{R}=-\nu'$, which can not hold for
all time. This is possible for all $x_{i}$ if $F'=R'=\nu'=0$.
This implies that at some $r$, $F'=f'=A'=C'=B'_{i}=0$
($i=1,2,...,n$). Hence we choose

\begin{equation}
\frac{F'}{(n+1)F}>-\nu' ~~~~\text{and}~~~~ \frac{R'}{R}>-\nu'
\end{equation}

Also from the solution (4) we have

$$
-\nu'=\frac{A'(r)\sum_{i=1}^{n}x_{i}^{2}+\sum_{i=1}^{n}B'_{i}(r)x_{i}+C'(r)}
{A(r)\sum_{i=1}^{n}x_{i}^{2}+\sum_{i=1}^{n}B_{i}(r)x_{i}+C(r)}
$$
Now writing in a quadratic equation in $x_{1}$ we have for real
$x_{1}$,

$$
\nu'^{2}+\left\{2(A'+A\nu')\sum_{k=2}^{n}x_{k}+\sum_{k=2}^{n}(B'_{k}+B_{k}\nu')
\right\}^{2}\le \sum_{i=1}^{n}B_{i}'^{2}-4A'C'
$$
So
$$
\nu'^{2}|_{max}=\sum_{i=1}^{n}B_{i}'^{2}-4A'C'
$$
Hence from (35) we have

\begin{equation}
\frac{F'}{(n+1)F}\ge
\sqrt{\sum_{i=1}^{n}B_{i}'^{2}-4A'C'}~,~~~\forall ~r
\end{equation}

which implies $F'\ge 0$, $\forall~r$. Now for $R'+R\nu'>0$, we
shall study the three possible choices separately.\\

(i)~ $f(r)=0$, $\Lambda=0$:\\

Here the solution for $R$ can be written as

$$
R^{\frac{n+1}{2}}=\frac{(n+1)}{2}\sqrt{F(r)}~(t-a(r))
$$

So as $t\rightarrow a$,~~
$R^{\frac{n-1}{2}}R'+R^{\frac{n+1}2{}}\nu'\rightarrow
-\sqrt{F(r)}~a'(r)$ and as $t\rightarrow\infty$,
~~$\frac{R'}{R}+\nu'\rightarrow \frac{F'}{(n+1)F}+\nu'$. Hence
for $R'+R\nu'>0$ we must have $a'<0$ and
$\frac{F'}{(n+1)F}>\sqrt{\sum_{i=1}^{n}B_{i}'^{2}-4A'C'}$~.\\
\\

(ii)~ $f(r)=0$, $\Lambda=0$:\\

The solution for $R$ can be written as
$$
R^{\frac{n+1}{2}}=\sqrt{\frac{n(n+1)F(r)}{2\Lambda}}~Sinh\left[\sqrt{\frac{(n+1)\Lambda}
{2n}}~(t-a(r)\right]
$$

In this case as $t\rightarrow
a$,~~$R^{\frac{n-1}{2}}R'+R^{\frac{n+1}2{}}\nu'\rightarrow
-\sqrt{F(r)}~a'(r)$ and as $t\rightarrow\infty$,
~~$\frac{R'}{R}+\nu'\rightarrow
\frac{F'}{(n+1)F}-\sqrt{\frac{2\Lambda F}{n(n+1)}}~a'(r) +\nu'$.
Thus for $R'+R\nu'>0$ we must have $a'(r)<0$ and
$\frac{F'}{(n+1)F}-\sqrt{\frac{2\Lambda
F}{n(n+1)}}~a'(r)>\sqrt{\sum_{i=1}^{n}B_{i}'^{2}-4A'C'}~.$\\
\\

(iii)~~$f(r)\neq 0, \Lambda=0,~ \dot{R}(t_{i},r)=0,~n=3$:\\

Here the solution for $R$ is
$$
R^{2}=r^{2}-\frac{F(r)}{r^{2}}(t-t_{i})^{2}~.
$$

The limiting value of $\frac{R'}{R}+\nu'$ ~as~$t\rightarrow
\infty$ will be $\frac{F'}{2F}+\nu'-\frac{1}{r}$. Hence for
$R'+R\nu'>0$ we should have $\frac{F'}{2F}+\frac{1}{r}>
\sqrt{\sum_{i=1}^{n}B_{i}'^{2}-4A'C'}$~.

\section{{\protect\normalsize \textbf{Discussion}}}

In the last two sections a details study of shell crossing
singularity has been done for dust model with or without
cosmological constant for Szekeres model of ($n+2$)-D space-time.
The physical conditions for shell crossing singularity are
presented in section III. These conditions however do not depend
on $\Lambda$ (whether zero or not) and the form of the conditions
are identical for the three cases presented there. For
geometrical conditions the locus of shell crossing depends on the
discriminant of the co-ordinate variables $x_{i}$'s
($i=1,2,...,n$). If both $\frac{R'}{R}$ and $\frac{F'}{(n+1)F}$
are greater than $\sqrt{\sum B'^{2}_{i}-4A'C'}$ then there will be
no shell crossing singularity even if $R'+R\nu'=0$. Here $\rho$
is finite and $R'+R\nu'=0$ will correspond to a real extrema for
$\beta$. On the other hand if $\frac{R'}{R}=\sqrt{\sum
B'^{2}_{i}-4A'C'}$ then shell crossing singularity is a 2-surface
in $(n+2)$-dimensional space-time. For $\frac{R'}{R}<\sqrt{\sum
B'^{2}_{i}-4A'C'}$, the shell crossing set lie on a
$n$-hypersphere and it intersects with constant $(t,~r)$ along
the curve $\frac{R'}{R}=-\nu'=$ constant. For future work it will
be interesting to study in details the possibility of shell
crossing singularity with pressure.\\\\

{\bf Appendix: A detailed study of a quadratic expression:}\\

Consider a general quadratic expression in $n$ variables

\begin{equation}
z=a\sum^{n}_{i=1}x_{i}^{2}+\sum^{n}_{i=1}b_{i}x_{i}+c
\end{equation}

For the present problem (given in equation (31)) we have

\begin{equation}
a=R'A-RA'~,~~b_{i}=R'B_{i}-RB'_{i}~,~~c=CR'-C'R
\end{equation}

Equation (37) can be rewritten as

\begin{equation}
z=a\sum^{n}_{i=1}\left(x_{i}+\frac{b_{i}}{2a}
\right)^{2}+\frac{d}{4a}~,~~d=4ac-\sum^{n}_{i=1}b_{i}^{2}
\end{equation}

By substitution from (38) we obtain

\begin{equation}
d=MR^{2}+NR'^{2}+LRR'
\end{equation}

with

$$
M=4A'C'-\sum^{n}_{i=1}B_{i}^{'2}~,~~N=4AC-\sum^{n}_{i=1}B_{i}^{2}~,~~L=-4(A'C+AC')+2\sum^{n}_{i=1}B_{i}B'_{i}
$$

Using equation (6) one sees that $N=1$ and $L=0$ and $d$
simplifies to

$$
d=MR^{2}+R'^{2}
$$

We shall now discuss the sign of $z$ for the following cases:\\

$(i)~~ a>0, ~d>0:$ It is clear from equation (39) that $z$ is
positive definite for all values of $x_{i}$'s.\\

$(ii)~~ a>0, ~d=0:$ In this case $z\ge 0$. The equality sign
occurs for a particular value of the variables $x_{i}$'s namely
$x_{i}=-\frac{b_{i}}{2a}~,~i=1,2,...,n$.\\

$(iii)~~ a>0, ~d<0:$ Here $z$ has indefinite sign. In particular,
$z$ will zero when the variables will lie on a hypersphere having
centre
$(-\frac{b_{1}}{2a},-\frac{b_{2}}{2a},...,-\frac{b_{n}}{2a})$ and
radius $\frac{\sqrt{|d|}}{~~2a}$.\\

If $a<0$ we have indefinite sign of $z$ for $d<0$ and for $d>0$,
$z$ will be negative definite (which is not possible in the
present paper). We note that if $z=0$ (with $d<0$) then as above
the variables $x_{i}$'s lie on the hypersphere having the same
centre and radius is $\frac{\sqrt{|d|}}{~~2a}$. But if the
variables lie inside the above hypersphere then $z$ will have
positive value.\\\\

\textbf{Acknowledgement:}\newline

One  of  the  authors (S.C.)  is  thankful  to  IUCAA, Pune
(where a part of the  work  has been  done) for  worm  hospitality
and facilities for  this  work. Also UD is thankful to CSIR,
Govt. of India for providing research project grant (No.
25(0153)/06/EMR-II). \\

\textbf{References:}\newline
\newline
$[1]$ H. J. Seifert, {\it Gen. Rel. Grav.} {\bf 10} 1065 (1979).\\
$[2]$ C. Hellaby and K. Lake, {\it Astrophys. J.} {\bf 282} 1
(1984); {\it Astrophys. J.} {\bf 290} 381 (1985).\\
$[3]$ K. Lake, {\it Phys. Rev. D} {\bf 29} 771 (1984); A.
Benerjee, A. Sil and S. Chatterjee, {\it Astrophys. J.} {\bf 422}
681 (1994).\\
$[4]$ A Papapetrou and Hamoui, {\it Ann. Inst. Henri Poincare
Sect. A} {\bf VI} 343 (1967).\\
$[5]$ D. J. R. Swatton and C. J. S. Clarke, {\it Class. Quantum
Grav.} {\bf 15} 2891 (1998).\\
$[6]$ C. J. S. Clarke and N. O'Donnell, {\it Rend. Sem. Mat.
Univ. Politec. Torino.} {\bf 50} 39 (1992).\\
$[7]$ S. M. C. V. Goncalves, {\it Phys. Rev. D} {\bf 63} 124017
(2001).\\
$[8]$ B. C. Nolan, {\it Phys. Rev. D} {\bf 60} 024014 (1999); {\it
Class. Quantum Grav.} {\bf 20} 575 (2003).\\
$[9]$ C. Hellaby and A. Krasinsky, {\it Phys. Rev. D} {\bf 66}
084011 (2002).\\
$[10]$ S. Chakraborty and U. Debnath, {\it Int. J. Mod. Phys. D} {\bf 13} 1085 (2004); {\it gr-qc}/0304072.\\
$[11]$ U. Debnath, S. Chakraborty and J. D. Barrow, {\it Gen.
Rel. Grav.} {\bf 36} 231 (2004); {\it gr-qc}/0305075.\\

\end{document}